\documentclass[aps,prl,twocolumn,groupedaddress]{revtex4}
\usepackage{graphicx}
\usepackage{amsmath}
\usepackage[pstricks,mediumqspace,squaren]{SIunits}

\begin{document}
\title{Stabilizing a purely dipolar quantum gas against collapse}
\author{%
  T. Koch\footnotemark[1]\footnotetext[1]{t.koch@physik.uni-stuttgart.de (TK).},
  T. Lahaye, J. Metz,\\
  B. Fr\"ohlich, A. Griesmaier,
  T. Pfau\footnotemark[1]\footnotetext[1]{t.pfau@physik.uni-stuttgart.de (TP).}
}
\affiliation{%
  \centerline{%
    5. Physikalisches Institut, Universit\"at Stuttgart, %
    Pfaffenwaldring 57, 70550 Stuttgart, Germany.} %
} %

\begin{abstract}
  We report on the experimental observation of the dipolar collapse
  of a quantum gas which sets in when we reduce the contact
  interaction below some critical value using a Feshbach resonance.
  Due to the anisotropy of the dipole-dipole interaction, the
  stability of a dipolar Bose-Einstein condensate depends not only
  on the strength of the contact interaction, but also on the
  trapping geometry. We investigate the stability diagram and find
  good agreement with a universal stability threshold arising from a
  simple theoretical model. Using a pancake-shaped trap with the
  dipoles oriented along the short axis of the trap, we are able to
  tune the scattering length to zero, stabilizing a purely dipolar
  quantum gas.
\end{abstract}

\maketitle

Interactions between atoms dominate most of the properties of
quantum degenerate gases~\cite{dalfovo99}. In the ultracold regime
these interactions are usually well described by an effective
isotropic zero-range potential. The strength and sign of this
\emph{contact interaction} is determined by a single parameter,
the scattering length $a$. The contact interaction is responsible
for a variety of striking properties of quantum gases. Strongly
influencing the excitation spectrum of the condensate it gives
rise to e.g. the superfluidity of Bose-Einstein condensates (BEC)
or the existence of vortex lattices. The contact interaction also
plays a crucial role in the physics of strongly correlated systems
like in the BEC-BCS crossover~\cite{zwierlein05} or in quantum
phase transitions like the Mott insulator
transition~\cite{greiner02}.

Another fundamental topic is the question of the existence of a
stable ground state depending on the modulus and sign of the
contact interaction. In the homogeneous case repulsive contact
interaction ($a>0$) is  necessary for the stability of the BEC. In
contrast, if the contact interaction is attractive ($a<0$), the
BEC is unstable. This instability can be prevented by an external
trapping potential. The tendency to shrink towards the center of
the trap is in that case counteracted by the repulsive quantum
pressure arising from the Heisenberg uncertainty relation.
Detailed analysis~\cite{ruprecht95} yields that a condensate is
stable as long as the number of atoms  $N$ in the condensate stays
below a critical value $N_{\rm crit}$ given by
\begin{equation} \label{eq:acritcont}
N_{\rm crit}=\frac{k a_{\rm ho}}{|a|}
\end{equation}
where $a_{\rm ho}$ is the harmonic oscillator length and $k$ is a
constant on the order of $1/2$. This scaling, as well as the
collapse dynamics for $N>N_{\rm crit}$, have been studied
experimentally with condensates of
$^7$Li~\cite{sackett98,gerton00} and
$^{85}$Rb~\cite{donley01,roberts01}.
In~\cite{modugno02,ospelkaus06} the atom number dependance of the
collapse of mixtures of bosonic $^{87}$Rb and fermionic $^{40}$K
quantum gases has been investigated.

Being \emph{anisotropic} and \emph{long-range}, the dipole-dipole
interaction (DDI) differs fundamentally from the contact
interaction. Besides many other properties, the stability
condition therefore changes in a system with a DDI present.
Considering the case of a purely dipolar condensate with
homogeneous density polarized by an external field, one finds that
due to the anisotropy of the DDI, the BEC is unstable, independent
of how small the dipole moment is~\cite{goral00}. As in the pure
contact case a trap helps to stabilize the system. In the dipolar
case, however, it is not only the quantum pressure that prevents
the collapse but more importantly the anisotropic density
distribution imprinted by the trap. Consider a cylindrically
symmetric harmonic trap
\begin{equation}
V_{\rm trap}(r,z)=\frac{1}{2}m\left(\omega_r^2 r^2+\omega_z^2
z^2\right)
\end{equation}
with the dipoles oriented along $z$ and $r$ being the distance
from the symmetry axis. As can be seen in Fig.~1A, in a
pancake-shaped trap (aspect ratio $\lambda=\omega_z/\omega_r>1$)
the dipoles predominantly repel each other and the BEC is stable.
In contrast, a cigar-shaped trap ($\lambda<1$, Fig.~1B) leads to
mainly attractive forces and hence to a dipolar collapse.
Following this simple argument one expects that in the prolate
case a positive scattering length $a$ is needed to stabilize the
BEC, whilst in the oblate case one can even afford a slightly
negative $a$. The dependance of the stability of a dipolar BEC on
the trap aspect ratio $\lambda$ and scattering length $a$ has been
extensively studied theoretically~\cite{santos00,yi01,eberlein05},
and is experimentally investigated in this paper.

\begin{figure}
\begin{center}
\label{gra_trapgeo}
\includegraphics[width=7cm]{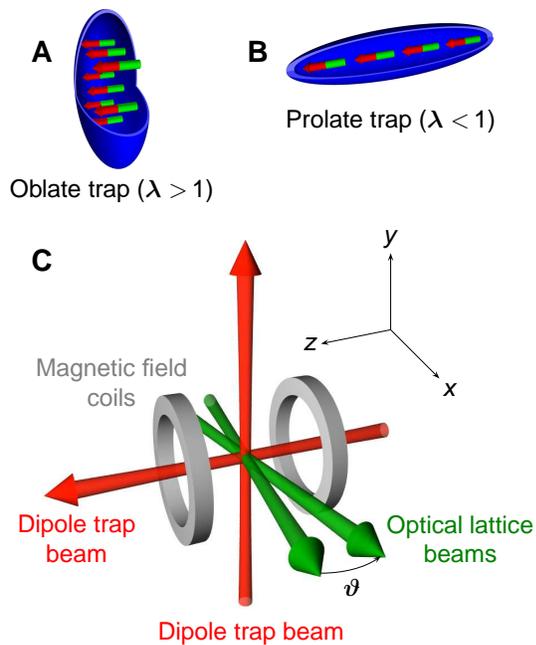}% Here is how to import EPS art
\caption{(A, B) Intuitive picture of the trap geometry dependance
of the BEC stability: In an oblate trap the dipoles mainly repel
each other, whereas in a prolate trap the interaction is
predominantly attractive. (C) The different trapping geometries
are realized by the crossed optical dipole trap (red) and an
additional 1D optical lattice (green). The magnetic field is
pointing along the symmetry axis $z$ of our traps.}
\end{center}
\end{figure}

Our measurements are performed with a BEC of
$^{52}$Cr~\cite{griesmaier05} which is to date the only
experimentally accessible quantum gas with observable
dipole-dipole interaction~\cite{stuhler05,lahaye07}. To compare
contact and dipolar interactions we introduce the length scale of
the magnetic DDI
\begin{equation}
a_{\rm dd}=\frac{\mu_0 \mu^2 m}{12 \pi \hbar^2}.
\end{equation}
The numerical prefactors in $a_{\rm dd}$ are chosen such that a
homogeneous condensate becomes unstable to local density
perturbations for $a\leq a_{\rm dd}$~\cite{santos03}. As Chromium
has a magnetic dipole moment of $\mu=6\mu_B$ ($\mu_B$ the Bohr
magneton), $a_{\rm dd}\simeq 15 a_0$, where $a_0$ is the Bohr
radius. Far from Feshbach resonances, $a$ takes its background
value $a_{\rm bg}\simeq 100 a_0$~\cite{werner05} and the BEC is
stable for any $\lambda$. To explore the unstable regime we thus
reduce the scattering length, which in the vicinity of a Feshbach
resonance scales like
\begin{equation}
a=a_{\rm bg}\left(1-\frac{\Delta B}{B-B_0}\right)\ \label{eq:a}
\end{equation}
with the applied magnetic field $B$. To be able to tune $a$
accurately we use the broadest of the resonances in
$^{52}$Cr~\cite{werner05} which is located at
$B_0\simeq\unit{589}{G}$ and has a width of $\Delta
B\simeq\unit{1.5}{G}$~\cite{lahaye07}.

\begin{figure}
\begin{center}
\includegraphics[width=7cm]{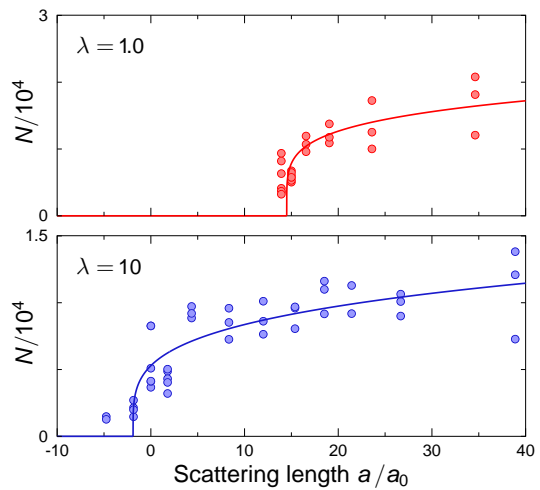}% Here is how to import EPS art
\caption{\label{gra_Nvsa}The atom number $N$ in the condensate as
a function of $a$ for two traps having different aspect ratios
$\lambda$. The solid lines are fits to Eq.~\ref{eq:powerlaw} used
to determine the critical scattering length $a_{\rm crit}$ (see
text).}
\end{center}
\end{figure}

The details of our experimental setup and procedure have already
been described elsewhere~\cite{griesmaier07,lahaye07} and shall
only be summarized here. We produce a BEC of approximately 25,000
atoms about $\unit{10}{G}$ above the resonance where the
scattering length is still close to its background value. Once the
BEC is obtained, we adiabatically shape the trapping potential to
the desired aspect ratio $\lambda$ within $\unit{25}{ms}$. In
order to be able to vary $\lambda$ over a large range, we generate
the trapping potential by a crossed optical dipole trap and a
superimposed one-dimensional optical lattice along the
$z$-direction (see Fig.~1C). The two lattice beams (wavelength
$\lambda_{\rm latt}=\unit{1064}{nm}$, waist $w_{\rm latt}=
\unit{110}{\mu m}$, maximum power per beam $P_{\rm latt}=
\unit{5}{W}$) propagate in the $xz$-plane under a small angle of
$\vartheta/2=4^\circ$ with respect to the $x$-axis. This
configuration creates a standing wave along the $z$-axis with a
spacing $d=\lambda_{\rm latt}/[2 \sin(\vartheta
/2)]=\unit{7.6}{\mu m}$. Due to the large spacing we load at most
two sites when ramping up the optical lattice. Tunneling processes
are completely negligible on the timescale of our experiments. By
varying the powers in the beams we are able to provide nearly
cylindrically symmetric traps, with aspect ratios $\lambda$
between $\sim 1/10$ and $\sim 10$, while keeping the average trap
frequency $\bar{\omega}=(\omega_r^2 \omega_z)^{1/3}$ approximately
constant~\cite{footnote_lambda}.

We then ramp the magnetic field within $\unit{10}{ms}$ to adjust
the value of the scattering length. The current providing the
magnetic field is actively stabilized on the $10^{-5}$
level~\cite{footnote_noise}, which results in a resolution of
$\Delta a\sim a_0$ around the zero crossing of the scattering
length. After an additional holding time of $\unit{2}{ms}$ we
finally switch off the trap and take an absorption image along the
$x$-axis, after a time of flight of $\unit{5}{ms}$. The BEC atom
number and radii are obtained by fitting the density profile using
a bimodal distribution~\cite{griesmaier07}. Knowing the atom
number and radii we can calibrate the scattering length $a$ as a
function of the magnetic field $B$~\cite{giovanazzi06,lahaye07}.

We observe two effects when approaching the zero-crossing of the
scattering length: The BEC shrinks in both directions due to the
decreasing scattering length and the ellipticity of the cloud
changes as a manifestation of the enhanced dipolar
effects~\cite{lahaye07}. Finally, when we decrease the scattering
length even further, the BEC atom number abruptly decreases. At
this point the density distribution does not show a bimodal shape
any more but becomes thermal-like. The total atom number stays
roughly constant during this collapse, excluding three-body loss
processes causing the decrease in BEC atom number.
The critical scattering length $a_{\rm crit}$ where the condensate
collapses depends strongly on the trap aspect ratio $\lambda$
(Fig.~2). For an isotropic trap (red) the collapse occurs at
$a\simeq15 a_0$, whereas the pancake-shaped trap (blue) can even
stabilize a purely dipolar BEC ($a\simeq0$).

\begin{table}
\begin{center}
\caption{\label{table_traps}Trap frequencies and aspect ratios of
the traps that we used. The trap frequencies were measured by
either exiting the center of mass motion or parametric heating and
are accurate to about ${10}{\%}$.}
\begin{tabular}{cccccc}
\hline \hline
Trap&$\omega_r/(2\pi)$ (Hz)&$\omega_z/(2\pi)$ (Hz)&$\bar{\omega}/(2\pi)$ (Hz)&$\lambda=\omega_z/\omega_r$\\
\hline
1&1300& 140&620& 0.11\\
2&890& 250 & 580&0.28\\
3& 480& 480&480& 1.0\\
4& 530 & 1400 & 730&2.6 \\
5& 400& 2400 & 730&6.0 \\
6& 330& 3400&720& 10\\
\hline \hline
\end{tabular}
\end{center}
\end{table}

We repeated this experiment for all the six traps listed in
Table~\ref{table_traps}, thereby covering a range of two orders of
magnitude in the trap aspect ratio $\lambda$.
By fitting to the observed BEC atom numbers~(Fig.~2) the threshold
function
\begin{equation}
\label{eq:powerlaw} N=\max\left[0,N_0(a-a_{\rm
crit})^\beta\right],
\end{equation}
where $N_0$, $a_{\rm crit}$ and $\beta$ are fitting parameters, we
find the critical scattering length $a_{\rm crit}$. The simple
functional form (Eq.~\ref{eq:powerlaw}) was chosen because it
accounts for the slowly decreasing BEC atom number when
approaching the collapse point. The exponent $\beta$ describing
the steepness of the collapse was found to be $\beta \simeq0.2$
for all traps. The obtained values of $a_{\rm crit}$ versus the
trap aspect ratio are plotted in Fig.~3~A. We observe a clear
shift towards smaller $a$ as $\lambda$ increases. For the most
oblate trap ($\lambda=10$) we can reduce the scattering length to
zero and hence access the purely dipolar regime experimentally.

\begin{figure*}
\begin{center}
\includegraphics[width=16cm]{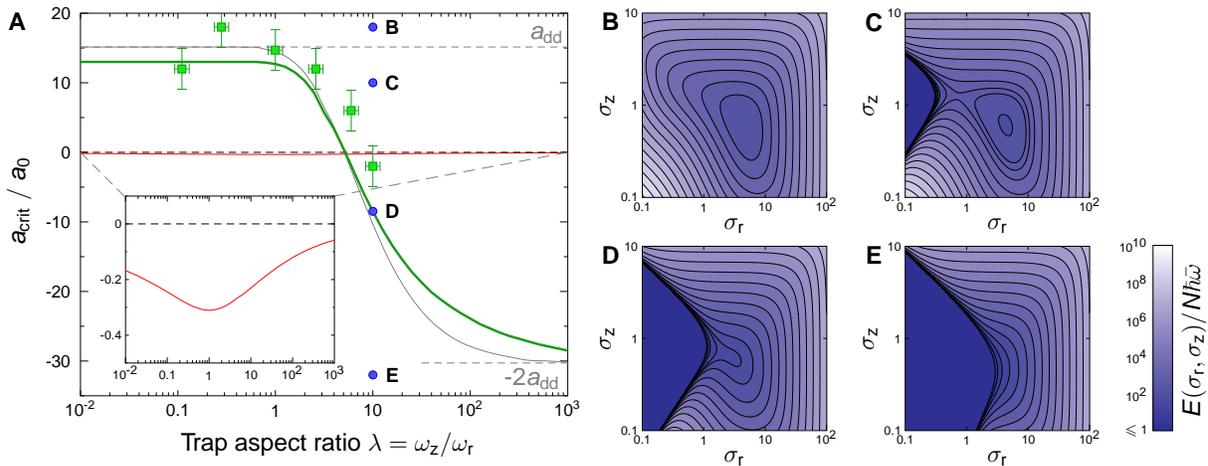}
\caption{\label{gra_acollvsAR} (A) Stability diagram of a dipolar
BEC. Experimental (green squares) and theoretical (green line)
values of the critical scattering length $a_{\rm crit}$ are
plotted as a function of the trap aspect ratio. The theory curve
is obtained for 20,000 atoms and an average trap frequency
$\bar{\omega}=2 \pi \times \unit{700}{Hz}$ (the average values we
find for our six traps). The red curve (magnified in the inset)
marks the stability threshold for a BEC with pure contact
interactions using the same parameters. In grey we plot the
asymptotic stability boundary ($N\rightarrow \infty$) which for
$\lambda\rightarrow 0$ ($\lambda\rightarrow \infty$) converges to
$a_{\rm dd}$ ($-2 a_{\rm dd}$), see text. (B-E) Behavior of the
energy landscape $E(\sigma_r,\sigma_z)$. Lines of equal energy are
plotted for fixed $\lambda=10$ and four different values of the
scattering length $a$ (blue dots in (A)). For $a_{\rm
crit}<a<a_{\rm dd}$ (C) the collapsed prolate ground state emerges
($\sigma_r\rightarrow0$ at finite $\sigma_z$) and the BEC becomes
metastable.}
\end{center}
\end{figure*}
To get a more quantitative insight into the collapse threshold
$a_{\rm crit}(\lambda)$ we numerically determine the critical
scattering length (green curve in Fig.~3~A) as follows. The ground
state wave function $\Phi({\boldsymbol r})$ of a BEC can be found
by the minimization of the Gross-Pitaevskii energy
functional~\cite{dalfovo99}
\begin{widetext}
\begin{equation}
E[\Phi]=\int\left[\frac{\hbar^2}{2m}|\nabla \Phi|^2+ V_{\rm
trap}|\Phi|^2+ \frac{g}{2} |\Phi|^4+\frac{1}{2}|\Phi| ^2\int
U_{\rm dd}({\boldsymbol r}-{\boldsymbol r}') |\Phi({\boldsymbol
r}')|^2{\rm d}{\boldsymbol r}' \right]{\rm d}{\boldsymbol r}.
\label{eq:energf}
\end{equation}
\end{widetext}
Here the first term corresponds to the kinetic energy $E_{\rm
kin}$, the second to the potential energy $E_{\rm pot}$ in the
trap, while the third represents the contact interaction energy
$E_{\rm cont}$, where $g=4 \pi \hbar^2 a / m$ is the coupling
constant. The last, non-local term $E_{\rm dd}$ arises from the
magnetic DDI~\cite{giovanazzi03}, where
\begin{equation}
U_{\rm dd}({\boldsymbol r})=\frac{\mu_0\mu^2} {4
\pi}\frac{1-3\cos^2\theta}{|{\boldsymbol r}|^3}
\end{equation}
is the interaction energy of two magnetic dipoles $\mu$ aligned by
an external field. Here ${\boldsymbol r}$ is the relative position
of the dipoles and $\theta$ the angle between ${\boldsymbol r}$
and the direction $z$ of polarization. In order to obtain an
estimate of $a_{\rm crit}$ we calculate the energy
$E(\sigma_r,\sigma_z)$ of a cylindrically symmetric Gaussian wave
function
\begin{equation}
\Phi(r,z)=\left(\frac{N}{\pi^{3/2}\sigma_r^2 \sigma_z a_{\rm
ho}^3}\right)^{1/2}\exp\left(-\frac{1}{2 a_{\rm ho}^2}\left(
\frac{r^2}{\sigma_r^2}+\frac{z^2}{\sigma_z^2}\right)\right)
\end{equation}
with $\sigma_r$ and $\sigma_z$ as variational parameters. Using
this ansatz, where $a_{\rm ho}=\sqrt{\hbar/(m \bar{\omega})}$, the
contributions to the total energy are the zero point fluctuations
\begin{equation}\label{eq:E_kin}
\frac{E_{\rm kin}}{N \hbar\bar{\omega}}=\frac{1}{4}
\left(\frac{2}{\sigma_r^2}+\frac{1}{\sigma_z^2}\right),
\end{equation}
the potential energy
\begin{equation}\label{eq:E_pot}
\frac{E_{\rm pot}}{N \hbar\bar{\omega}}=\frac{1}{4
\lambda^{2/3}}\left(2
\sigma_r^2+\lambda^2 \sigma_z^2\right),
\end{equation}
and the mean-field interaction energy
\begin{equation}\label{eq:E_int}
\frac{E_{\rm cont}+E_{\rm dd}}{N \hbar\bar{\omega}}=\frac{N a_{\rm
dd}}{\sqrt{2\pi}a_{\rm ho}}\frac{1}{\sigma_r^2
\sigma_z}\left(\frac{a}{a_{\rm dd}}-f(\kappa)\right).
\end{equation}
Here $f(\kappa)$ is a monotonically decreasing function of the
condensate aspect ratio $\kappa=\sigma_r/\sigma_z$ with the
asymptotic values $f(0)=1$ and $f(\infty)=-2$, arising from the
non-local term in Eq.~\ref{eq:energf}~\cite{giovanazzi03}. The
function $f$ vanishes for $\kappa=1$ implying that for an
isotropic density distribution the magnetic DDI does not
contribute to the total energy. To obtain $a_{\rm crit}$ we lower
the scattering length until the energy landscape
$E(\sigma_r,\sigma_z)$ does not contain a minimum for finite
$\sigma_r$ and $\sigma_z$ any more (see Fig.~3~B-E). Starting with
large values $a>a_{\rm dd}$ we find that $E(\sigma_r,\sigma_z)$
supports a global minimum for finite $\sigma_r$ and $\sigma_z$
independently of $\lambda$ and thus the BEC is stable (Fig~3~B).
Going below $a\sim a_{\rm dd}$ the absolute ground state is a
collapsed infinitely thin cigar-shaped BEC
($\sigma_r\rightarrow0$) and the possible existence of an
additional local minimum (corresponding to a metastable state) is
determined by the trap aspect ratio $\lambda$ (see Fig~3~C, where
$a_{\rm dd}>a>a_{\rm crit}$ and Fig~D, where $a=a_{\rm crit}$).
Finally, below $a\sim -2 a_{\rm dd}$ (Fig~3~E) the metastable
state vanishes for any $\lambda$ and the BEC is always
unstable~\cite{yi01,eberlein05}.

Considering the limit $N a_{\rm dd}/a_{\rm ho}\gg1$ where the
terms (\ref{eq:E_kin}) and (\ref{eq:E_pot}) can be
neglected~\cite{footnote_Nadd/aho} (grey curve in Fig.~3~A) we
gain further insight into the nature of the dipolar collapse. In
this case the stability is governed by the competition between the
contact and dipole-dipole interaction only, that is by the sign of
the last term in Eq.~\ref{eq:E_int}. Hence the critical scattering
length is implicitly given by
\begin{equation}\label{eq:acritdd}
a_{\rm crit}(\lambda)=a_{\rm dd} f\left( \kappa \left( \lambda
\right) \right).
\end{equation}
The asymptotic behavior of the theory curve $a_{\rm crit}=a_{\rm
dd}$ for $\lambda\rightarrow 0$ (respectively $a_{\rm crit}=-2
a_{\rm dd}$ for $\lambda\rightarrow\infty$) now becomes apparent
as for extremely prolate (respectively oblate) traps the cloud
shape follows the trap geometry and $f$ takes on its asymptotic
values. Another particular point is $a_{\rm crit}=0$ marking the
aspect ratio $\lambda$ needed to stabilize a purely dipolar BEC.
More precisely, as $f(1)=0$, we search for the trap in which the
ground state of a purely dipolar BEC is isotropic. As the DDI
tends to elongate the BEC along the $z$-direction and shrink it
radially, it is clear that the desired trap is oblate. Using our
model we obtain the criterion $\lambda>\lambda_{\rm c}\approx5.2$
for a purely dipolar BEC to be stable, a result that agrees well
with the values found in~\cite{yi01,baranov02,eberlein05}.

The grey curve in Fig.~3~A that we obtain by numerically solving
Eq.~\ref{eq:acritdd} shows a universal behavior in the sense that
in the large $N$ limit $a_{\rm crit}(\lambda)$ does not depend any
more on the absolute values of the trap frequencies and $N$. This
fact clearly distinguishes the dipolar collapse from the pure
contact case (red curve in Fig.~3~A). The former is ruled by the
interplay between contact and dipolar interaction whereas in the
latter the zero point energy and the contact interaction rival
against each other. Due to the different $N$-scaling of the two
competing terms in the pure contact case, the
$\lambda$-dependance, which is already weak for finite
$N$~\cite{gammal01}, completely vanishes in the limit of large $N$
as the stability criterion reads $a_{\rm crit}(\lambda)= 0$ (see
Eq.~\ref{eq:acritcont} and red curve in Fig.~3~A). Furthermore the
stability threshold obtained here applies for any dipolar system
like e.g. hetero-nuclear molecules, where the only difference is
the specific value of $a_{\rm dd}$.

In spite of the simplicity of our model we find good agreement
between experiment and theory (Fig.~3~A). We checked that the
different atom numbers and mean trap frequencies that we find for
the six traps modify the green curve by much less than the error
bars which arise mainly from the calibration of the scattering
length. For the  most oblate trap ($\lambda=10$) the
$1/e$-lifetime of the purely dipolar BEC ($a=0$) decreases to
$\sim \unit{13}{ms}$.

In conclusion, we experimentally mapped the stability diagram of a
dipolar BEC. The dependance on scattering length and trap aspect
ratio agrees well with a simple model based on the minimization of
the energy of a Gaussian ansatz. By using a pancake-shaped trap we
were able to enter the regime of purely dipolar quantum gases.
This work opens up the route to new and exciting
physics~\cite{baranov02}. A clear subject for future studies is
the dynamics of the dipolar collapse, which might show anisotropic
features. Another remarkable property of a dipolar BEC in a
pancake-shaped trap is the existence of a roton minimum in its
Bogoliubov spectrum~\cite{santos03}. Furthermore, close to the
collapse threshold, the existence of structured ground states is
predicted~\cite{ronen07,dutta07}, a precursor for the supersolid
phase~\cite{goral02a} that is expected to appear in dipolar BECs
in three dimensional optical lattices. Finally, a field that has
gained increasing interest in the recent past is the study of
unusual vortex lattice patterns in rotating dipolar
BECs~\cite{cooper05}.

We would like to thank L. Santos, G.~V. Shlyapnikov and
H.-P. B\"uchler for stimulating discussions and M. Fattori for his
contributions in earlier stages of the experiment. We acknowledge
financial support by the German Science Foundation (SFB/TR 21 and
SPP 1116) and the EU (Marie-Curie fellowship contract
MEIF-CT-2006-038959 to T.L.).


\begin{thebibliography}{10}
\bibitem{footnote_lambda}
  Larger aspect ratios having the same mean frequency could not be
  achieved. In this case the radial confinement is too weak to hold
  the atoms against gravity and residual magnetic field gradients.
\bibitem{footnote_noise}
  In the interesting range between $\unit{1}{Hz}$ and
  $\unit{10}{kHz}$ (faster noise is filtered out by eddy currents in
  the vacuum chamber) the relative current noise is smaller than
  $1\times 10^{-5}$ rms.
\bibitem{footnote_Nadd/aho}
  For our typical experimental parameters we have $N a_{\rm
    dd}/a_{\rm ho}\simeq30$.
\end{thebibliography}
\end{document}